# CO gas from the aftermath of a giant impact in the inner region of a planetary system

Tajana Schneiderman[1*], Luca Matrà[2], Alan P. Jackson[3,4], Grant M. Kennedy[5,6], Quentin Kral[7], Sebastián Marino[8], Karin I. Öberg[9], Kate Y. L. Su[10], David J. Wilner[9], Mark C. Wyatt[8]

**Models of terrestrial planet formation predict that the final stages of planetary assembly, lasting tens of millions of years beyond the dispersal of young protoplanetary disks, are dominated by planetary collisions. It is through these giant impacts that planets like the young Earth grow to their final mass and achieve long-term stable orbital configurations. A key prediction is that these impacts produce debris. To date, the most compelling observational evidence for post-impact debris comes from the planetary system around the nearby 23 Myr-old A star HD 172555. This system shows large amounts of fine dust with an unusually steep size distribution and atypical dust composition, previously attributed to either a hypervelocity impact or a massive asteroid belt. Here, we report the spectrally resolved detection of a CO gas ring co-orbiting with dusty debris between ~6-9 au - a region analogous to the outer terrestrial planet region of our Solar System. Taken together, the dust and CO detections favor a giant impact between large, volatile-**

[1] Department of Earth, Atmospheric and Planetary Sciences, Massachusetts Institute of Technology, Cambridge, Massachusetts, USA
[2] Centre for Astronomy, School of Physics, National University of Ireland Galway, University Road, Galway, Ireland
[3] Centre for Planetary Sciences, University of Toronto at Scarborough, 1265 Military Trail, Toronto, ON M1C 1A4, Canada
[4] School of Earth and Space Exploration, Arizona State University, 781 E. Terrace Mall, Tempe, AZ 85287, USA
[5] Department of Physics, University of Warwick, Coventry CV4 7AL, UK
[6] Centre for Exoplanets and Habitability, University of Warwick, Gibbet Hill Road, Coventry CV4 7AL, UK
[7] LESIA, Observatoire de Paris, Université PSL, CNRS, Sorbonne Université, Univ. Paris Diderot, Sorbonne Paris Cité, 5 place Jules Janssen, F-92195 Meudon, France
[8] Institute of Astronomy, University of Cambridge, Madingley Road, Cambridge CB3 0HA, UK
[9] Center for Astrophysics | Harvard & Smithsonian, 60 Garden Street, Cambridge, MA 02138, USA
[10] Steward Observatory, University of Arizona, 933 N. Cherry Avenue, Tucson, AZ 85721, USA

**rich bodies. This suggests that planetary-scale collisions, analogous to the Moon-forming impact, can release large amounts of gas as well as debris, and that this gas is observable, providing a window into the composition of young planets.**

HD 172555 is a 23±3 Myr-old[1] A type star with a mass of 1.76 $M_\odot$[2] and luminosity of 7.7 $L_\odot$, located 28.5 pc[3,4] from Earth within the young β Pictoris moving group[5]. Its planetary system hosts large amounts of dust in the terrestrial region, producing an infrared (IR) excess best fit by warm (290 K) dust[6] with an infrared luminosity $7.2\times10^{-4}$ times that of its host star[7]. Spatially resolved mid-IR observations constrain this dust to a disk of material in the inner, <10 au region, viewed close to edge-on from Earth[8,9]. Rare solid-state emission features detected at 8-9.3µm[10] indicate the presence of glassy silica (tektites and obsidian) and solid SiO. These require high temperature processing and vapor condensation, supporting a hypervelocity (>10 km/s) impact between planetary bodies for the origin of the dust[11,12]. Detailed, self-consistent modelling indicates that this impact scenario can explain the spectral features, as well as the overabundance of sub-µm sized grains and steep size distribution inferred from the *Spitzer* IR spectrum[12].

We used the Atacama Large Millimeter/submillimeter Array (ALMA) to detect dust continuum and CO J=2-1 gas emission from the HD 172555 system (Fig. 1). The compact, spatially unresolved dust and gas emission originates from within 15 au of the star, consistent with previous IR and optical observations. The spectrum of CO in Fig. 2a is extracted from the ALMA data cube and is spectrally resolved. The double-peaked profile is expected from gas orbiting in Keplerian rotation around the central star. The centroid of the CO profile is at a heliocentric velocity of 2.3±0.2 km/s, consistent with the stellar velocity[13], confirming that the gas emission is associated with the HD 172555 system.

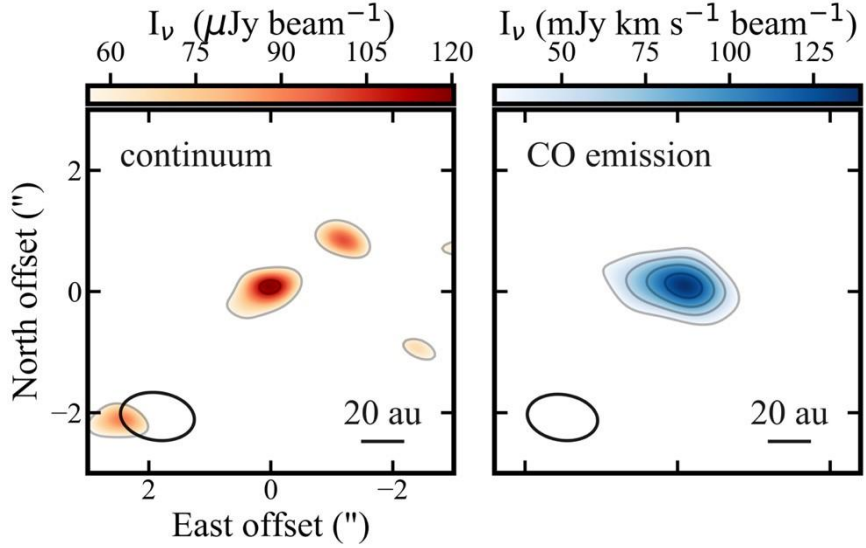

**Figure 1** | Cleaned emission maps of the HD 172555 system. Left panel is a ~4$\sigma$ detection of the dust continuum emission. Right panel is the moment-0 CO J=2-1 transition, with a ~9$\sigma$ detection. Contour levels are at 2$\sigma$ intervals, with $\sigma$ being 0.029 mJy beam$^{-1}$ and 15 mJy km s$^{-1}$ beam$^{-1}$ for the continuum and CO moment-0 maps, respectively. Beam size denoted in lower left corner of each panel.

We modelled the CO emission as a vertically thin ring of optically thin gas with radially and azimuthally uniform surface density between an inner and outer boundary, in Keplerian rotation around the central star. We find that the CO is confined to a ring of radius ~7.5 au and width ~3.3 au. The data is consistent with a symmetric disk model; we find no strong evidence of asymmetry between the blue- and red-shifted sides of the CO spectral line profile. We convert the integrated flux of 120 mJy km/s to a CO gas mass considering non-local thermodynamic equilibrium excitation, finding masses in the range between $(0.45-1.21) \times 10^{-5}$ M$_\oplus$ for temperatures between 100 and 250 K. The total millimetre dust mass, as calculated from the observed continuum emission, is $(1.8\pm0.6)\times10^{-4}$ M$_\oplus$ for an expected equilibrium blackbody temperature of 169 K at 7.5 au.

To test our assumption that the CO J=2-1 emission is optically thin, we create a full 3D radiative transfer model of the ring with the spatial morphology described above. For a gas temperature comparable to, or higher than the expected blackbody temperature of 169 K at 7.5 au, we find low optical depths, which validates our optically thin assumption. This corresponds to total CO masses of (1.4±0.3) ×$10^{-5}$ $M_\oplus$ at 169 K. On the other hand, we find that colder (<<100 K) temperatures would require higher optical depth, and much higher CO masses to reproduce our observed CO line emission (Fig. 2b). This temperature-mass degeneracy can be resolved by the detailed spectral shape of the emission since models of optically thin, warm gas constrained to a narrow ring produce sharper peaks compared to the optically thick, cold models (Fig. 2a). We find that the data prefers less massive, warmer models (lower $\chi^2$).

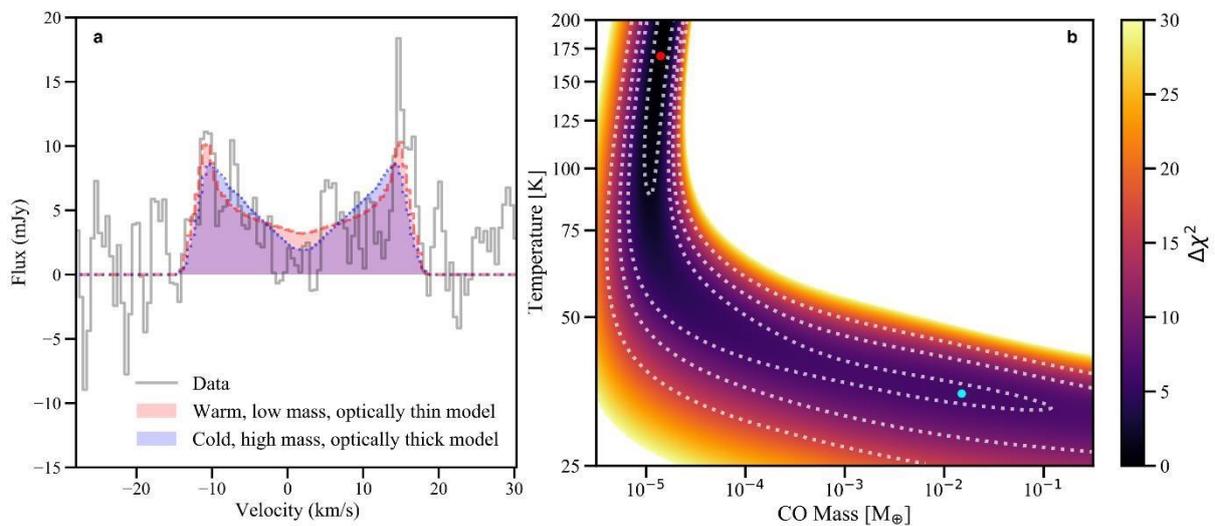

**Figure 2 | a.** CO J=2-1 spectrum of HD172555 (grey solid line), compared to the output of two radiative transfer models: a low mass, warm temperature CO model (red dashed line), and a high mass, cold temperature CO model (blue dotted line). **b.** $\chi^2$ map (where the $\chi^2$ is relative to the best-fit model) showing a clear CO temperature-mass degeneracy for models that are good fits to the data (darker on the colour scale). The red and light blue dots correspond to the red (warm) and blue (cold) CO models shown in panel a. Dotted contours enclose models that are consistent with the data at the >[1,2,3,4] σ confidence level.

The observed CO gas in the circumstellar environment around HD 172555 will be subject to photodissociation from the stellar and interstellar UV field. The former dominates at the 7.5 au location of the CO ring and causes CO (if unshielded) to be rapidly destroyed on a timescale of ~1 day. However, the lifetime of CO is extended when taking into account shielding effects from atomic carbon[14] (C, as produced by CO phodissociation), molecular hydrogen ($H_2$), and CO itself (self-shielding), that prevent UV photons from penetrating into the ring. Self-shielding alone, in the warm temperature, low CO mass scenario, extends the gas lifetime by ~2-4 years. Given the lack of observational constraints on C and $H_2$ column densities, we cannot definitively estimate the shielding and hence the CO destruction timescale. However, CO could have easily survived over the system age (23 Myr) both in a scenario where C and CO dominate the gas mass (C/CO ratio of > 0.16-0.43, see Methods), and in an $H_2$ dominated scenario ($H_2$/CO > ~$10^5$). Therefore, CO could either be the result of a very recent production and/or replenishment mechanism, or have survived for a large fraction of the system age.

The CO detection constrains the gas to be located in the ~6-9 au region of the HD172555 planetary system, co-located with the dust[9]. Accounting for the difference in luminosity between the Sun and HD172555, this region corresponds to the same thermal conditions as ~2.1-3.3 au - the Asteroid belt region - in our Solar System. This makes the presence of CO, a highly volatile gas with sublimation temperatures as low as ~20 K, extremely surprising in a system with age 23 Myr. Some physical mechanism must explain the presence of gas and dust in the outer terrestrial planet forming region of HD 172555. We test four hypotheses for the origin of the debris, and examine them in light of existing evidence from both dust observations and from the ALMA detection of CO in the system. These scenarios are: (i) leftover gas and dust from a primordial, protoplanetary disk, (ii) collisional production within an extrasolar asteroid belt, (iii) inward transport of material from an external reservoir, and (iv) release in the aftermath of a giant impact between planetary-sized bodies.

Young A-type stars are born surrounded by protoplanetary disks of primordial gas and dust, but only ~2-3% survive beyond the first 3 Myr of a star's lifetime[15,16]. Even if the CO observed around HD 172555 were primordial, with its lifetime extended through shielding, the system would remain a remarkable outlier not only in age (at 23 Myr old), but also in dust mass (orders of magnitude lower than protoplanetary disks[17]) and in radial distribution of CO and dust

(constrained within 10 au, and with a CO cavity, in contrast to protoplanetary disks typically extending out to tens/hundreds of au[18,19]). The extreme depletion in dust mass would require efficient dust removal, either through accretion onto the star or grain growth[17]. At the same time, the confined radial extent would require interior/exterior truncation, for example by as-yet undiscovered companions. Finally, HD172555's peculiar dust mineralogy (requiring energetic processing) is seldom seen in young protoplanetary disks, although it may have arisen in nebular shocks analogous to those that could have led to chondrule formation in the Solar System[20]. In conclusion, a primordial scenario would make HD172555 an extreme outlier amongst other protoplanetary disks, favouring instead second-generation production.

In the second-generation case, a steady-state collisional cascade within an asteroid belt can explain the dust mass detected in the system[21], but not its abnormally steep particle size distribution[12], or its mineralogy requiring high-energy collisions at velocities higher than expected within typical belts[10–12]. While CO gas is commonly observed within collisional cascades in colder extrasolar Kuiper belt analogs at tens of au[22], there its presence can be explained by release of CO gas initially trapped within icy bodies, or by desorption of $CO_2$ ice followed by rapid photodissociation of $CO_2$ gas[23–27]. This picture is not plausible at 7.5 au; studies of ice-bearing asteroids in the Solar System show that while retaining water ice in the deep subsurface is possible, explaining the presence of outgassing main-belt comets in the Asteroid belt, CO and $CO_2$ cannot remain trapped at these temperatures[28,29]. Additionally, any bodies forming at this location would be too warm to have formed with a significant reservoir of CO or $CO_2$.

Alternatively, the observed dust and CO might originate from inward scattering of small bodies from an outer, cold exocometary reservoir[30]. We first explore the possibility of gas and dust production through sublimation of Solar System -like, 10-km sized exocomets. We find that this scenario can only reproduce the observations if exocomets can be delivered on low eccentricity orbits, finding that the exocomet replenishment rate could potentially be reconciled with the non-detection of an outer belt (see Methods). However, this scenario cannot explain the relatively narrow radial distribution of the CO ring, or the mineralogy of the observed dust, and is thus not likely. On the other hand, it is possible that a single, cold, massive icy body is scattered inwards from an undetected outer belt. To produce the observed, largely axisymmetric distribution of material and dust mineralogy, the icy body would need to undergo a giant impact and release its CO or $CO_2$ content. Assuming a 25% CO+$CO_2$ ice mass fraction, and complete ice release at impact, we find that a dwarf planet (at least half the size of Pluto) would be needed to produce the (lower limit on the) CO gas mass observed.

Finally, CO gas and dust could be produced by a giant impact between planetary bodies formed in situ at ~7.5 au. The epoch of terrestrial planet formation, lasting from ~10-100 Myr, is expected to be dominated by giant impacts. Within the solar system, there is abundant evidence for the occurrence of giant impacts; the iron enrichment of Mercury[31,32], the formation of the Moon[33–35], and the Martian hemispheric dichotomy[36,37] are all hypothesized to have their origins in giant impacts. The HD 172555 system, at 23 Myr, is at the expected age that terrestrial planet formation proceeds through griant imapcts; the dust and gas observed in the system are located in a region that is analogous to the terrestrial zone in our own solar system. Studies of post-impact dynamics allow us to set constraints on progenitor masses and time since impact to be placed by considering the current spatial distribution and mass of dust (see Methods). The dust in the system is axisymmetric within observational uncertainties, implying that the time since impact is at least the debris symmetrization timescale, which at 7.5 au is of order ~0.2 Myr[38]. The width of the dust debris, as resolved at shorter wavelengths[8,9,i], suggests progenitor masses of order ~ 8 $M_\oplus$, though the exact value is dependent on the radial width of the dust debris, see Methods. Impacts between such planets produce debris that would survive encounters with leftover planets, on a timescale longer than symmetrization; this indicates that impacts of such planets could be responsible for the observed, long-lived debris field.

Further supporting this scenario, we find that the optically thin CO gas mass detected in the system ($(0.45-1.21) \times 10^{-5}$ $M_\oplus$), or ~10 times the mass of Earth's atmosphere, is consistent with post-impact release from a planetary atmosphere. We note that any $CO_2$ in a planetary atmosphere will rapidly be converted into observable CO if liberated from the atmosphere, as $CO_2$ cannot be significantly shielded by C in the same way as CO because its photodissociation bands extend further toward the optical[39]. Simulations show that up to 60% of a modestly-sized, heavy atmosphere can be stripped in the initial shock of an impact[40,41]. We find that the observed CO mass, as well as the C needed to shield the CO for at least the symmetrization timescale, require the release of an amount of $CO_2$ corresponding to just 9-23% of the total present in the Venusian atmosphere[42]. More (less) massive planets with similar heavy atmospheres would require a smaller (larger) fraction of the atmosphere removed, or a smaller (larger) abundance of CO and/or $CO_2$. For lighter, $H_2$-dominated atmospheres, longerterm thermal effects can result in the stripping of the entire atmospheric envelope[43], in which case once again an amount of CO and/or $CO_2$ consistent with the observations could plausibly be liberated.

The detection and morphology of CO gas, combined with previous evidence from dust imaging and spectroscopy, supports a picture where a giant impact took place at least 0.2 Myr ago in the outer terrestrial planet forming region of the 23 Myr-old HD 172555 system. Planetary-scale impacts are predicted to be commonplace in the latest stages of planet formation; the discovery of CO gas in the terrestrial planet forming region, in amounts consistent with the expectation from atmospheric stripping, suggests that giant impacts may not only release copious, observable dust, but also detectable amounts of gas. This highlights the importance of gas release in post-impact dynamics, and opens the possibility of using gas as a tool to search for giant impacts in nearby planetary systems, and as a unique window into the composition of young planets and their atmospheres.

# Methods

**ALMA observations.** We analyzed archival data from the ALMA telescope taken during Cycle 1 in band 6 (project code 2012.1.00437.S). The observations were performed with the 12-m array in a compact antenna configuration. The on-source time was 76 minutes. The spectral setup included four spectral windows, of which three were set up in time division mode for continuum observations and centered at 213, 215, and 228 GHz. One window was set in frequency division mode (with high spectral resolution of 488.29 kHz) to target the $^{12}$CO J=21 line (rest frequency 230.538 GHz). We calibrated the visibility data using scripts provided by the ALMA observatory. The CASA software[ii] (version 5.6.1) was used for visibility imaging. For both the continuum and CO line emission, we flagged data from antennas 7, 19, and 25. These data were taken in a hybrid configuration, with the three flagged antennas far from the compact group of antennas. Flagging these three antennas significantly improved the imaging.

The CLEAN algorithm[44] was used to image both the continuum and CO line emission through the CASA *tclean* task. To achieve maximum sensitivity, we used natural weighting for both datasets, resulting in a synthesized beam size of 1.16" x 0.75", corresponding to 32.9 x 21.3 au at the known distance to the system and position angle of 81□. Before line imaging, we subtracted the continuum (measured in line-free regions of the spectrum) in visibility space using the CASA *uvcontsub* task. We then imaged the CO to produce a cube with frequencies within ± 25 MHz of the rest frequency, corresponding to velocities within ± 30 km/s of the radial velocity of the star. The data was imaged at the native spectral resolution (twice the original channel width) of 488.29 kHz. To obtain the CO moment-0 map, shown in Figure 1, we integrated along the velocity axis between ± 15 km/s of the stellar velocity.

The emission is spatially unresolved in both the continuum and CO images, which have RMS noise levels of 0.029 mJy beam$^{-1}$ and 26 mJy km s$^{-1}$ beam$^{-1}$, respectively. This yields a peak detection at a SNR per beam of 4 and 6 for the continuum and CO J=2-1 spectrally integrated emission, and total fluxes of 0.12 ± 0.03 mJy and 170 ± 30 mJy km/s, respectively. Note that the flux calibration is expected to be accurate within 10%; this uncertainty was added in quadrature to obtain the quoted errors[iii]. Both the CO gas and continuum dust emission centroids are consistent with the *Gaia* proper-motion corrected position of the system[3,4]. To extract a 1D spectrum from the data cube, we integrate within a circular mask of radius 2.5$\sigma_{beam}$ (with $\sigma_{beam}$ being the standard deviation of the 2D Gaussian beam, averaged between the major and minor axis).

**Millimetre dust mass.** To calculate the dust mass we assume the dust grains act as blackbodies and emit according to their Planck function at a temperature of 169 K. Assuming the dust is optically thin, the total dust mass can be calculated from the observed emission ($F_\nu = 0.12 \pm 0.03$ mJy), when accounting for the stellar contribution (0.035 mJy at 230 GHz from a Rayleigh-Jeans extrapolation of IR measurements). The dust grain opacity is assumed to be 10 cm$^2$ g$^{-1}$ at 1000 GHz and scaled to the frequency of the observation with an opacity power law index of $\beta = 1$ [45]. These assumptions yield a dust mass of $(1.8 \pm 0.6) \times 10^{-4}$ M$_\oplus$.

**Optically thin CO ring modelling.** To model the velocity spectrum expected from a circular orbiting ring or disk of material, we calculate Keplerian velocities assuming a stellar mass of 1.76M$_\odot$[2]. 2D orbital velocity vectors are calculated for a radial and azimuthal grid, assuming a vertically thin ring/disk of gas that has radially and azimuthally uniform surface density between an inner and outer boundary. They are transformed to the sky plane using the ring/disk inclination to obtain radial velocities along the line-of-sight. These velocities in the reference frame of the star are then added to the radial velocity of the star in the barycentric frame (left as a free parameter) to obtain barycentric velocities as observed by ALMA. A histogram of these velocities, with the same binning as the observed data, serves as a model spectrum. We rescale the unitless spectrum by a factor linked to the integrated flux of the line, which we leave as a free parameter in the fit, to obtain a spectrum in the same units as the data. This spectrum is then convolved with a Gaussian of FWHM equal to twice the channel width to reproduce the spectral response of the instrument due to Hanning smoothing[iv]. In addition to these parameters, the model fits the inclination, radial location of the midpoint, and width of the ring/disk.

A Markov Chain Monte Carlo (MCMC) approach was used to determine the best fit to the data. We used the Python package *emcee*[46]. The uncertainty in each velocity bin was assumed to be equal to the RMS measured in the region of the spectrum outside the detected emission. Flat priors were applied for the radial location, width, integrated line flux, and stellar velocity. A Gaussian prior was applied to the inclination, assuming the gas shares the same inclination as the dust disk, as determined from previous resolved imaging[9]. We carried out an additional run where a flat prior was applied to the inclination, to confirm the assumption of shared inclination. Extended Data Fig. 1 shows the posterior probability distributions of the parameters obtained from our MCMC runs, whereas Extended Data Table 1 indicates the best-fit values, obtained as the 50$^{th}$ ± 34$^{th}$ percentiles of the posterior distributions of each parameter, marginalized over all other parameters. Best fit values from both runs assuming a Gaussian or flat prior on the inclination are included.

**Optically thin CO mass calculation.** To derive a CO mass from the best-fit spectrally-integrated line flux, we begin by assuming that the line is optically thin and considering the excitation conditions the gas may be subject to, which affect this conversion. We follow an existing framework[23], which considers that the energy levels of a CO molecule may be populated by collisions with other species (or by one dominant species, the main collisional partner), or by absorption and emission of radiation, giving rise to two limiting regimes, a radiation-dominated regime (at low gas densities), and a collision-dominated regime (local thermodynamic equilibrium, LTE, at high gas densities). The choice or density of collisional partners (in our case, electrons) does not affect the level populations in these two limiting regimes, and therefore the range of CO masses derived.

To account for the full range of excitation conditions, we therefore use a non-LTE code[23] to solve the statistical equilibrium equations and calculate the level populations. This includes the effect of fluorescence induced by stellar UV and IR radiation as seen by a CO molecule at 7.5 au from the central star[47]. For the star, we adopt a PHOENIX[48] model spectrum fitted to multiband optical and near-IR photometry using the MultiNest code[49], as described in previous works[50,51]. This yields a best-fit stellar effective temperature of $T_{\text{eff}} = 7840 \pm 30$ K, and luminosity of $L_\star = 7.7 \pm 0.1$ $L_\odot$. The non-LTE calculation yields a formal range of possible CO masses between $(0.45 - 1.21) \times 10^{-5}$ $M_\oplus$ for kinetic temperatures between 100 and 250 K, encompassing the blackbody temperature of 169 K at 7.5 au.

**3D radiative transfer modelling.** We use the RADMC-3D[11,v] radiative transfer code to check the impact of optical depth in more detail. We use the same ring geometry as obtained from optically thin fitting (Extended Data Table 1, Gaussian prior). We vary the input CO mass and kinetic temperature over a 2D grid (Figure 2b), and connect the latter to the vertical aspect ratio by assuming a vertically isothermal gas disk with a mean molecular weight of 14 (i.e. the gas mass is dominated by atomic C and O, as expected in a second generation scenario). In order to sample the ring well spatially and spectrally, we create cubes of J=2-1 emission with a pixel size of 6 mas (corresponding to a physical scale of 0.18 au), and the same native channel width as our data. We then spectrally convolve with a Gaussian to reproduce the spectral response of the instrument and extract a 1D model spectrum by spatially integrating the model emission. We then compared this model spectrum to the data (as shown in Figure 2a) and calculated a $\chi^2$ for every mass and temperature in our grid, to obtain a $\chi^2$ map (Figure 2b).

---

[11] http://www.ita.uni-heidelberg.de/~dullemond/software/radmc-3d/

**CO survival lifetime against photodissociation.** The observed CO gas in the circumstellar environment around HD 172555 will be subject to photodissociation from the stellar and interstellar UV field. The photodissociation rate in s$^{-1}$ of a molecule in a radiation field I($\lambda$) is $k = \int \sigma(\lambda)I(\lambda)d\lambda$, where $\sigma(\lambda)$ is the photodissociation cross section in cm$^2$. We use the CO photodissociation crosssections from the Leiden database[39,vi]. We adopt the stellar spectrum from the optically thin mass section above, scaled to the center of the gas ring (7.5 au) to obtain the stellar radiation field. We find that the star dominates over the interstellar radiation field, and that the CO photodissociation timescale (1/k) at the ring's radial location is approximately a day. The shielding effects are estimated using pre-computed shielding constants[39]; we interpolate the constants for stars of 4000 and 10000 K to a stellar temperature of 8000 K, closest to HD172555's effective temperature. The CO column density is calculated from the center of the ring along the line of sight to the star, using our best-fit uniform ring model parameters. To find the H$_2$ and C column densities that provide sufficient shielding, we interpolate the shielding constants along the column density axis and find the column density required.

**Delivery from an outer belt.**

*Replenishment requirement* We consider a scenario where the dust grains and CO gas are produced from sublimation-driven release by Solar System -like comets entering the inner region of the HD172555 system. This requires replenishment of the observed total dust mass (in grains up to cm in size) on timescales comparable to their removal (assuming steady state). We assume their removal is dominated by collisions, setting up a cascade down to the smallest grains (of size ~3.5 μm for a grain density of 2700 kg m$^{-3}$) that are then removed by radiation pressure from the central star, to derive a mass loss rate of 2.2 x 10$^{-2}$ M$_\oplus$ Myr$^{-1}$ (using Eq. 21 in [25]). We neglect the effect of gas drag, assuming that the latter leaves larger, cm grains unaffected. We use the results of thermochemical modelling[52] calibrated on Solar System comets to estimate the mass loss rate per unit surface area of an exocomet at 7.5 au around a 7.7 L$_\odot$ to be 5.6 × 10$^{-6}$ kg m$^{-2}$ s$^{-1}$ of dust in grains up to cm in size. Dividing by an assumed bulk density of 560 kg m$^{-3}$, and assuming a 1:1 dust/ice ratio, this corresponds to an erosion rate of 0.62 m yr$^{-1}$. Therefore, a comet with a 10 km radius will be emitting dust at a rate of 7040 kg/s (3.72 x $10^{-8}$ $M_\oplus$Myr$^{-1}$), and will survive (if continuously sublimating at this rate) for ~ 16 kyr.

*High eccentricity exocomet population* In the first case, we assume exocomets arise from a yet undetected belt at 100 au (a typical location of cold exocometary belts around A stars[53]) and approach the inner regions on eccentric orbits. The velocity distribution from the observed CO emission implies line-of-sight velocities for the gas of ~14 km/s, corresponding to 7.5 au for circular orbits around HD172555. However, these velocities will be achieved at larger radii for comets with non-zero eccentricities, random pericentre directions, and orbiting in the same plane as HD172555's inner system. The pericentre distance corresponding to pericentre velocities of ~14 km/s will increase with the apocenter (and therefore the eccentricity) of the orbits; for an apocenter of 100 au, we derive a pericentre distance of 13.5 au. Therefore, we expect gas released from exocomets on these orbits to be located at ~13.5 au from the central star. For comparison, this is 80% larger than the radial location derived for circular orbits (7.5 au). However, eccentric exocomets with the observed velocities would produce a CO ring with diameter ~27 au (0.95"), which is comparable to the resolution element of our observations, implying the ring would have been marginally resolved, which is not the case. Additionally, scattered light observations strongly constrain dust emission to <0.9" (3σ) diameter, with a sharp outer edge. Therefore, these larger radii make an eccentric exocomet scenario inconsistent with the available observations.

*Low eccentricity exocomet population* In the second case, we assume exocomets are being continuously scattered inward from an outer belt by a chain of low mass planets, undergoing multiple scatterings and producing a low eccentricity comet population at ~7.5 au. We here assume circular orbits for simplicity. The dust observed requires replenishment at a rate of 2.2 x $10^{-2}$ $M_\oplus$Myr$^{-1}$. Assuming exocomets of 10 km in size with an exocometary dust release rate of 3.72 x $10^{-8}$ $M_\oplus$Myr$^{-1}$, 5.9 x $10^5$ exocomets are required to be sublimating at around ~7.5 au at any point in time. For a bulk density of 560 kg m$^{-3}$, this corresponds to 2.3 x $10^{-4}$ $M_\oplus$ in 10 km exocomets. While sublimating at this rate, such an exocomet would survive for ~16 kyr, so comets would need to be resupplied by inward scattering to the inner planetary system at a rate of 1.4 x $10^{-8}$ $M_\oplus$yr$^{-1}$. Inward scattering is an inefficient process; simulations maximising inward scattering by chains of low mass planets indicate that only a few % of comets encountering an outermost planets make it into the inner regions, as the vast majority are ejected[54]. Therefore, higher supply rates of order ~$10^{-6}$ $M_\oplus yr^{-1}$ from a putative outer belt are likely needed, which would imply that a currently undetected outer belt would have resupplied 23 $M_\oplus$ in 10 km exocomets into the inner regions over the 23 Myr age of the system.

We can compare this to the upper limit on the presence of an outer belt at 100 au from our ALMA data. Assuming blackbody temperatures, and the same dust opacity as used for dust in the inner regions, we derive an upper limit of $< 1.8 \times 10^{-3}$ $M_\oplus$ (3σ) on the mass of solids of sizes up to cm sizes. Extrapolating from cm sized grains up to 10 km exocomets (assuming a size distribution with a constant power law slope of -3.5) we obtain an upper limit on the total mass in 10 km sized exocomets of $< 1.8$ $M_\oplus$ (3σ). Therefore, the *current* mass of the outer belt would be at least a factor ~10 smaller than the mass that has been removed from the belt over its lifetime, which is by itself not impossible.

However, it is likely that exocomets reaching their inner regions would retain some eccentricity, which would prolong their survival (lower sublimation rates), but also potentially make the gas velocity distribution inconsistent with the observations. In addition to that, the narrow CO radial distribution is in tension with production rates of CO as a function of heliocentric distance in sublimating Solar System comets[55], which show no clear enhancement in production rates at radii whose thermal conditions are comparable to HD172555's. While inward scattered, sublimating exocomets could produce a steeper than collisional size distribution of grains, down to sizes smaller than the radiation pressure blow-out limit, they would not be able to reproduce the peculiar dust mineralogy indicative of energetic processing typical of hypervelocity impacts, unless exocometary impacts in the inner regions contribute, and/or are the source of the dust and gas observed. In conclusion, sublimating exocomets could be replenished in the inner system at high enough rates to explain the observed dust emission, but are unlikely to explain the relatively narrow distribution of CO, and the dust mineralogy observed around HD172555.

**In situ giant impact constraints from axisymmetry and width of debris.** Observations of both CO and dust[8,9] are consistent with the CO and dust distribution being axisymmetric. Because of this, the time since impact must be at least the symmetrization timescale, which is on the order of a few tens of thousands of orbits[38] (~0.2 Myr at 7.5 au). Constraints on the planet mass can be derived from the width of the debris, as a proxy for the velocity dispersion of released material. The width, $dr$, is given by $dr = 2re_p$, where $e_p$ is the proper eccentricity of the orbiting debris[56]. This eccentricity is related to the velocity dispersion, $\sigma_v$, through $\sigma_v \sim \sqrt{1.5} e_p v_k$, where $v_k$ is the keplerian velocity at the given semimajor axis. We assume the velocity dispersion is related to the escape velocity of the colliding bodies[38] by $\sigma_v \sim 0.46 v_{esc} \sim 0.46 \sqrt{\frac{2GM_p}{R_p}}$. Thus, for an observed debris width $dr$, a planet of mass $M_p \sim 103 \frac{M_*^{3/2}}{\rho_p^{1/2}} \left(\frac{dr}{r^{3/2}}\right)^3$ in $M_\oplus$ is expected, where $\rho_p$ is the planet's bulk density in g cm$^{-3}$, $M_*$ is the stellar mass in $M_\odot$, and both r and dr are in au. If the planetary bodies colliding have rocky, Earth-like compositions, they will have bulk densities of ~5.5 g cm$^{-3}$. We assume that the solid debris is confined to the same radial width as the dust width derived from mid-IR Q band imaging[8], which finds $dr_{dust}/r \sim 1.2$. These assumptions yield a planetary mass on the order of ~8 $M_\oplus$. We note that mid-IR imaging is sensitive to small grains whose width might have been broadened due to radiation pressure from the central star; if radiation pressure inflated the width compared to that expected from the velocity dispersion, the planetary mass involved would be reduced. If instead the bulk planetary density is lower, the mass of planet involved in the collision would be larger.

**Constraints from debris survival.** Progenitor mass constraints must also be considered in the context of debris survival; the observed debris must survive encounters with a leftover planet in the time since impact. We conservatively assume that the mass of the largest product of the impact is on the order of the mass of the impact progenitors, and consider the outcome of encounters between debris and a surviving planet[57]. Given an assumed planetary density similar to that of the Earth, and a semimajor axis of 7.5 au, we first note that for a leftover planet with mass below ~1.8 $M_\oplus$, the predominant (eventual) outcome of an encounter will be accretion, whereas for masses above that threshold, the predominant (eventual) outcome will be ejection. However, timescales of accretion and ejection play a role in debris survival. These timescales must be at least as long as the symmetrization timescale (0.2 Myr) to ensure debris produced in an impact could be seen in the symmetrical structure observed today. For progenitor masses in the range derived from the debris width constraints above, the reaccretion and ejection timescales are much longer than the symmetrization timescale. Therefore, giant impact debris at 7.5 au is expected to survive encounters with leftover planets with masses on the order of 8 $M_\oplus$, as large as those of the progenitors, on a timescale longer than the symmetrization timescale.

**Acknowledgements.** We are grateful to John Biersteker for discussions on the liberation of atmospheres in the aftermath of giant impacts. This paper makes use of ALMA data ADS/JAO.ALMA\#2012.1.00437.S. ALMA is a partnership of ESO (representing its member states), NSF (USA) and NINS (Japan), together with NRC (Canada), NSC and ASIAA (Taiwan), and KASI (Republic of Korea), in cooperation with the Republic of Chile. The Joint ALMA Observatory is operated by ESO, AUI/NRAO and NAOJ. KIÖ acknowledges support from the Simons Foundation (SCOL \#321183).

**Author Contributions** T.S. led the optically thin modelling and discussion. L.M. led the radiative transfer modelling. Both authors were involved in data reduction, processing, and writing of the manuscript. All authors contributed to discussions of the results and commented on the manuscript.

**Competing interests declaration** The authors declare no competing interests.

**Data Availability** The ALMA program number for the presented data is 2012.1.00437.S and data can be found in the online ALMA archive.

**Code Availability** RADMC-3D (https://github.com/dullemond/radmc3d-2.0) and *emcee* (https://emcee.readthedocs.io/en/stable/) are available online. Requests for additional code should be addressed to T.S. (tajana@mit.edu).

**Additional Information** Correspondence and requests for materials should be addressed to T.S. (tajana@mit.edu).

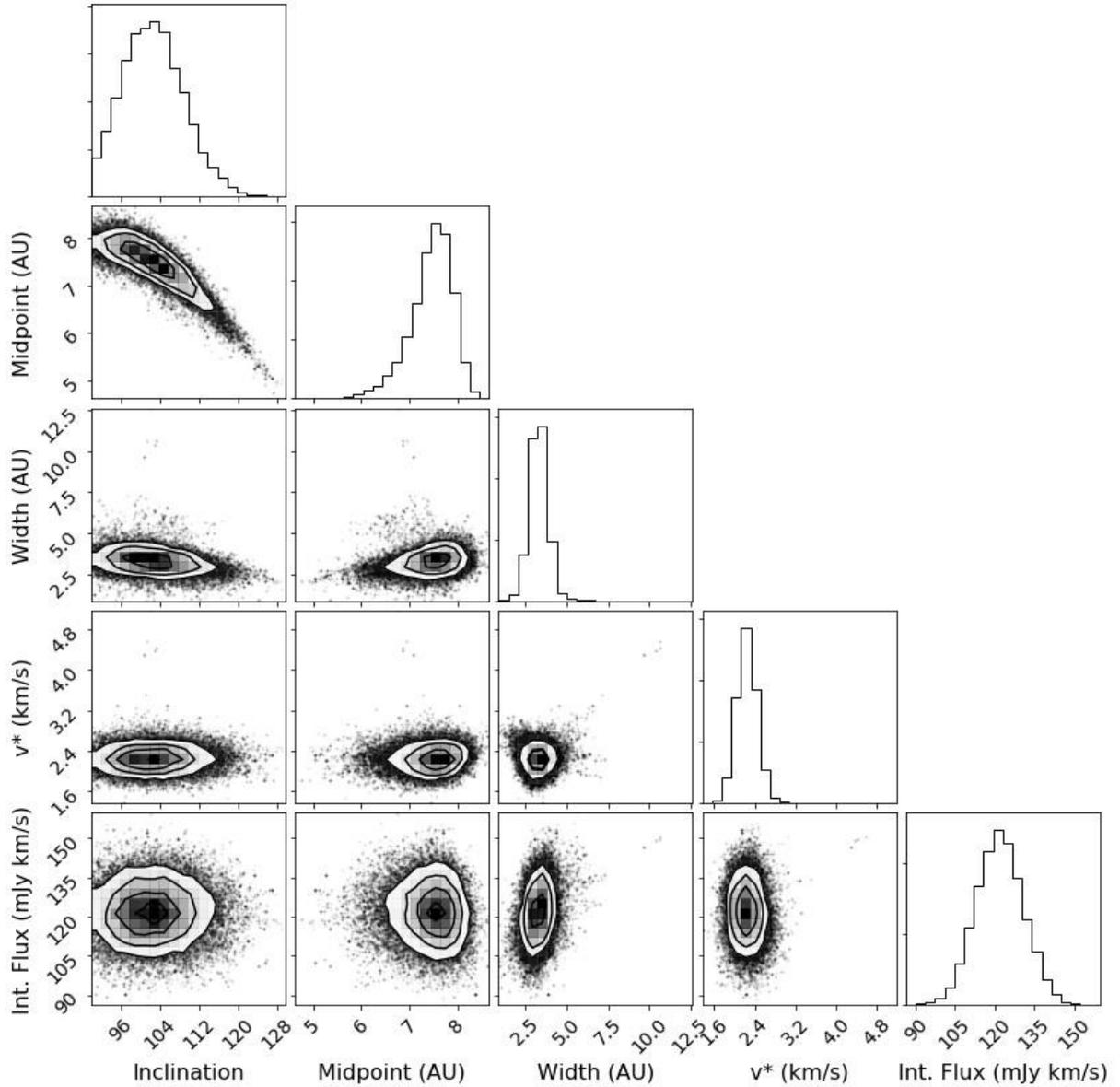

**Extended Figure 1.** Posterior probability distributions for the model parameters obtained from the emc*ee* fitting process. All parameters are well constrained, with best fit values listed in Extended Data Table 1. This model was fitted to spectral data retaining original channel widths, assumed a Gaussian prior on the inclination, and assumed a stellar mass of 1.76 M$_\odot$

| | Best fit parameters | |
|---|---|---|
| | Gaussian | Flat |
| Inclination (°) | 102 +6.0 -6.5 | 107 +12.1 -17.6 |
| Midpoint (au) | 7.4 +0.5 -0.4 | 7.4 +1.8 -0.6 |
| Width (au) | 3.4±0.5 | 3.1 +0.9 -0.7 |
| $V_*$ (km/s) | 2.3±0.2 | 2.3±0.2 |
| Int. Flux (mJy km/s) | 122 +8.9 -9.0 | 122 +8.5 -8.8 |

**Extended Data Table 1**: Best fit values (50±34 percentile) to the optically thin model of gas emission. Left column indicates values derived from the MCMC run where a Gaussian prior was applied to the inclination. Right column indicates values derived from the MCMC run where flat priors were applied to all model parameters.

[i] The literature[8] values for the dust location in au have been corrected to account for the new GAIA distance to the star compared to previous Hipparcos distance.
[ii] https://casa.nrao.edu/casadocs/casa-5.6.0
[iii] https://almascience.nrao.edu/documents-and-tools/cycle-1/alma-ot-reference-manual [iv] https://safe.nrao.edu/wiki/pub/Main/ALMAWindowFunctions/Note_on_Spectral_Response.pdf [v] http://www.ita.uni-heidelberg.de/~dullemond/software/radmc-3d/
[vi] https://home.strw.leidenuniv.nl/~ewine/photo/display_co_42983b05e2f2cc22822e30beb7bdd668.html